\documentclass[runningsheads]{llncs}
\usepackage{cite}
\usepackage{amsmath,amssymb,amsfonts}
\usepackage{algorithmic}
\usepackage{graphicx}
\usepackage{textcomp}
\usepackage[caption=false, font=normalsize, labelfont=sf, textfont=sf]{subfig}
 \usepackage{array,longtable}
\usepackage{booktabs,makecell,tabularx}
\usepackage{colortbl}
\usepackage{multirow}
\usepackage{xcolor}
\usepackage{xspace}
\usepackage{fancyhdr}
\usepackage{enumerate}
\begin{document}


\title{Longitudinal Compliance Analysis of Android Applications with Privacy Policies}

\titlerunning{Longitudinal Compliance Analysis of Apps}

\author{Saad Sajid Hashmi\inst{1} \and
Nazar Waheed\inst{2} \and
Gioacchino Tangari\inst{1} \and 
Muhammad Ikram\inst{1} \and 
Stephen Smith\inst{1}
}
\authorrunning{Hashmi et al.}
%
\institute{Macquarie University, Australia \\ \email{\{fname.lname\}@mq.edu.au} \and
University of Technology Sydney, Australia \\
\email{\{fname.lname\}@student.uts.edu.au}\\
}

\sloppy

\maketitle              
\begin{abstract}
Contemporary mobile applications (apps) are designed to track, use, and share users' data, often without their consent, which results in potential privacy and transparency issues. To investigate whether mobile apps have always been (non-)transparent regarding how they collect information about users, we perform a longitudinal analysis of the historical versions of 268 Android apps. These apps comprise 5,240 app releases or versions between 2008 and 2016. We detect inconsistencies between apps' behaviors and the stated use of data collection in privacy policies to reveal compliance issues. We utilize machine learning techniques for the classification of the privacy policy text to identify the purported practices that collect and/or share users' personal information, such as phone numbers and email addresses. We then uncover the data leaks of an app through static and dynamic analysis. Over time, our results show a steady increase in the number of apps' data collection practices that are undisclosed in the privacy policies. This behavior is particularly troubling since privacy policy is the primary tool for describing the app's privacy protection practices. We find that newer versions of the apps are likely to be more non-compliant than their preceding versions. The discrepancies between the purported and the actual data practices show that privacy policies are often incoherent with the apps' behaviors, thus defying the `notice and choice' principle when users install apps.

\end{abstract}



\section{Introduction}
``Privacy is the claim of individuals, groups, or institutions to determine for themselves when, how, and to what extent information about them is communicated to others.''\footnote{Alan Westin, Privacy and Freedom, 1967.} Between October 2012 and February 2013, Snapchat's privacy policy said, ``We do not ask for, track, or access any location-specific information from your device at any time while you are using the Snapchat application''~\cite{complaintFTC}. However Snapchat did the opposite by collecting and sharing user's geo-location information (Wi-Fi and cell-based location data) to its analytics tracking service provider.  
Federal Trade Commission (FTC) of the United States initiated an investigation, and in December 2014 ordered Snapchat to implement a comprehensive privacy program addressing risks to users' privacy~\cite{decisionFTC}. 

In February 2019, FTC fined TikTok 5.7 million USD for illegally collecting children's personal data~\cite{tiktok_ftc}. Between 2019-20, in the United States, several federal lawsuits against TikTok were filed citing the harvesting of users' (including children's) personal data without consent. In July 2020, these lawsuits were incorporated into a single class-action lawsuit. In February 2021, a settlement was reached where TikTok agreed to pay 92 million USD and stop the collection of users' bio-metric and location data~\cite{tiktok_pay}. Currently, TikTok is facing a privacy lawsuit in the United Kingdom for violating child privacy laws in the collection of personal information and sharing it with third parties~\cite{tiktok_uk}.

The examples of Snapchat, TikTok, and various other apps~\cite{children_apps} highlight that online service providers frequently do not comply with their privacy policies, despite the presence of regulations related to the disclosure of privacy practices. These regulations include the California Online Privacy Protection Act (CalOPPA)~\cite{caloppa}, California Consumer Privacy Act (CCPA)~\cite{ccpa}, and Children's Online Privacy Protection Act (COPPA)~\cite{coppa} in the US, General Data Protection Regulation (GDPR) in the EU~\cite{gdpr}, Data Protection Act in the UK~\cite{dpa}, and the Privacy Act in Australia~\cite{aus_privacy}. Although there are limitations to privacy policies (such as they are not often read~\cite{jensen2005privacy}, take a long time to read~\cite{mcdonald2008cost}, and are hard to comprehend~\cite{jensen2004privacy}), privacy policies remain the legally recognized standard of protecting privacy based on the ``notice and choice'' principle~\cite{cranor2012necessary}. This principle gives users a notice of the data practices performed by the app (or service) before that data is collected. The users then have the choice to {\it opt-in} (i.e., give permission) or {\it opt-out} (i.e., decline permission) of giving access to their details.

Several studies have shown the behavior of contemporary apps to be non-compliant with the stated disclosures in their privacy policies~\cite{slavin2016, wang2018, reyes2018won, okoyomon2019ridiculousness, han2019price, zimmeckEtAlMAPS2019, andow2020}. Given that the compliance landscape is continuously evolving, we aim to determine whether the non-compliant behavior has changed with time. In this study, we investigate the compliance of Android apps with their privacy policies over time, ranging  
from 2008 and 2016. 
We detect and analyze the personally identifiable information (PII) leaked by an app in disregard of the practices made public by the publisher in the app's privacy policy. We leverage machine learning (ML) techniques to classify the text of apps' privacy policies and to identify the purported practices that collect and/or share data (e.g., phone number). We then uncover the actual data leaks of an app through static analysis ({\it examining the app code}) and dynamic analysis ({\it inspecting the network traffic generated by the app}). While relying on existing techniques, this study is, to our knowledge, the first effort to integrate app and privacy policy text analyses to measure how the apps' privacy conducts have evolved over time. 

The main contributions of this paper are as follows:
\begin{itemize}
\item We identify the data collection practices ({\it privacy practices}) disclosed in 3,151 privacy policies (obtained from 2012 to 2019) of 405\footnote{{\it see} \S \ref{sec:apps} for clarification on different number of apps reported.} apps using machine learning classifiers. We find that 2,422 privacy policies from 327 different apps disclose at least one of these practices, with an average of 7.86 per policy (\S \ref{subsec:polAnalysis}).
\item We analyze historical versions of popular apps from 2008 to 2016, based on the union of leaks observed via static and dynamic analysis. 
The results show an increase in the average number of leaks per app version over time. Surprisingly, the average number of leaks to third-parties rises from 2.7 in 2011 to 4.43 in 2016 (\S \ref{subsec:leaksAnalysis}).
\item Our analysis reveals that the compliance of apps with their privacy policy is steadily decreasing from 33.16\% in 2011 to 10.76\% in 2016. Also, newer versions of the apps are performing more privacy policy violations than their immediately preceding version. For instance, 9.2\% of the app versions released in 2016 show an increase in first-party violations compared to their preceding version, whereas only 2.5\% show a decrease (\S \ref{subsec:compAnalysis}).     
\end{itemize}

The rest of the paper is organized as follows: In Section \ref{sec:dmethodology}, we describe our data collection approach and present our analysis methodology. In Section \ref{sec:analysis}, we discuss the findings of our analysis. Section \ref{sec:rwork} reviews the related work, and we conclude and discuss future research directions in Section \ref{sec:cfwork}.

\vspace{-0.2cm}
\section{Dataset and Methodology}
\label{sec:dmethodology}
\begin{figure}
    \centering
    \includegraphics[scale=0.4, keepaspectratio]{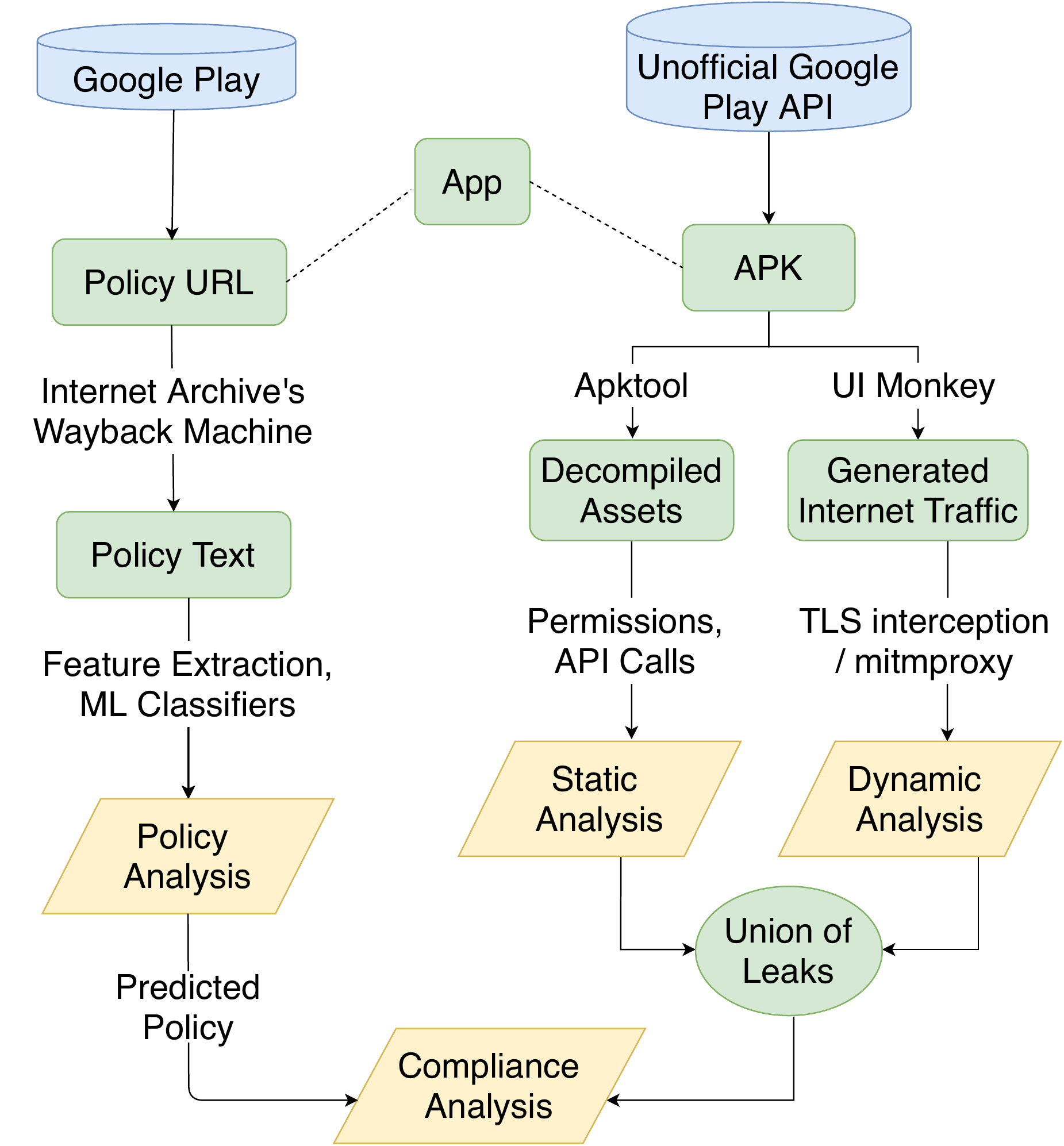}
    \caption{{\small Overview of our approach. We perform compliance analysis after taking the union of personal identifiable information (PII) leaks detected in both the static and dynamic analysis. }}
    \label{fig:bigPic}
\vspace{-0.3cm}
\end{figure}

Figure~\ref{fig:bigPic} depicts our methodology. In the following, we explain the steps involved in our data collection and analysis.

\subsection{Dataset}

\subsubsection{Mobile Apps}
We select Android apps that are \textit{(i)} popular, \textit{(ii)} have multiple versions, \textit{(iii)} are susceptible to network traffic interception, and \textit{(iv)} have privacy policy in the app's home-page on Google Play. The popular apps we select were either in the top 600 free apps based on the Google Play Store ranking, or in the top 50 in each app category, as of 10 Jan 2017. We discard apps that have less than four versions, as well as apps for which we can not intercept TLS (Transport Layer Security) traffic (e.g., apps with non-native Android TLS libraries such as Twitter, Dropbox). We also discard apps that do not contain privacy policy on their homepage at Google Play. For the compliance analysis, we collect 5,240 unique APKs that correspond to 268 apps.

We download the selected apps and their previous versions from an unofficial Google Play API~\cite{googlePlayAPI} that requires the package name and version code to download an APK. We infer the release date of the package from the last modification time of the files inside the APK (such as \texttt{AndroidManifest.xml} and \texttt{META-INF/MANIFEST.MF}). If the date is obviously incorrect (e.g. before 2008), we refer to third party services (appcensus.io, apkpure.com) for inferring the release dates.  

\vspace{-0.2cm}
\subsubsection{Privacy Policies}
Google requires Android app developers to disclose the collection and sharing of users' data~\cite{googlePlayDisclosure}. Accordingly, on Google Play Store, each application must provide a link to the privacy policy on its home-page. 

For the most updated version of each app, first, we access the URL of the privacy policy from the app's home-page ($https://play.google.com/store/apps/details?id=<pname>$ followed by the package name i.e., {\it pname}), and then scrape privacy policy link to extract the privacy policy text.   

To obtain the privacy policies of the previous versions of the same app, we leverage the Internet Archive's Wayback Machine~\cite{wayback}, which gives us access to previous snapshots of the app's privacy policy page. Since 1996, the Wayback Machine has archived full websites, including JavaScript codes, style sheets, and any multimedia resources that it can identify statically from the content of the website. We refer to the single capture of a web-page as a {\it snapshot}. Wayback Machine mirrors past snapshots of these websites on its servers. Note that we are unable to extract the privacy policy content if the URL obtained from the app's home-page refers to an invalid URL, or if the privacy policy is not found on that URL, or if the privacy policy web-page contains another link that refers to the privacy policy text. Also, the Wayback Machine poses some challenges (such as web-page redirect or varied frequency of archived snapshots
) in capturing a specific web-page~\cite{hashmi2019LCN}. This results in a miss to capture the privacy policy snapshot in that three-month window, and we move to the next three-month window.

There are instances where a single privacy policy is mapped to multiple app versions. This is because privacy policies only became mandatory on Google Play since 2018~\cite{google_gdpr}. Therefore, a privacy policy residing on a given URL today may not have been residing on the same URL a few years ago. If the Wayback Machine does not archive a policy's web-page, then the current policy's web-page (obtained in 2019) is mapped to all the app releases. Furthermore, we are unable to obtain policy web-pages prior to 2012. The app releases prior to 2012, therefore have a high difference in time with the privacy policy date. It may also be possible that the source code of an app is modified when releasing a new version however comes with the old privacy policy.

We use Memento API~\cite{momento} to capture snapshots at intervals of three months and obtain multiple privacy policies for the same app. Memento API provides the nearest time-stamp for the archived snapshot of a website from the date provided. By comparing the app version's release date with the snapshot dates of the privacy policy web-page, we obtain the snapshot that is immediately after the release date of the app version. For example, if an app released versions in Feb'16, Aug'16 and Dec'16, and the snapshots of the app's privacy policy are from Jan'16, Jul'16, and Jan'17, then the Feb'16 app version will be assigned the Jul'16 privacy policy. Similarly, Aug'16 and Dec'16 app versions will be assigned the Jan'17 privacy policy snapshot.

\vspace{-0.3cm}
\subsection{Privacy Policy Analysis}
\label{subsec:pp_analysis}

We characterize the identification of privacy practices (i.e., app behaviors that leak personally identifiable information (PII)) from the app's privacy policy as a supervised classification problem. We initially pre-process the privacy policy text, then we apply machine-learning classifiers to predict the privacy practices.


\vspace{-0.2cm}
\subsubsection{Data Pre-processing:}
From the URL of privacy policy, we obtain the text of the privacy policy by scraping the $<body>$ tag of the HTML page. Within the $<body>$ tag, we discard unnecessary content nested in other tags (such as $<script>$, $<style>$, $<meta>$, and $<noscript>$). Some archived pages in Wayback Machine have text in a particular format (for example, beginning with ``success'' and ending with ``TIMESTAMPS'') prefixed to the original text. We discard all prefix text prior to the analysis of privacy policy text.

\textit{Next}, we convert the privacy policy text into segments (or paragraphs) by ensuring that a single paragraph does not contain less than fifty characters, except if it is the last paragraph of the privacy policy. Previous work has shown that practices are better captured from the policy text if the text is broken down into segments~\cite{wilson2016ACL,Liu2018CMU}. If a text segment (other than the last segment) has less than fifty characters, it is likely to be the heading of the next paragraph; therefore, we merge it with the next text segment. We also combine two adjacent text segments if their combined length is less than 250 characters. After converting the text into segments, we lowercase all segments, and normalize whitespaces and apostrophes (for example, we change words like \textit{haven't} and \textit{don't} to \textit{have not} and \textit{do not} respectively). We finally discard non-alpha and non-ascii characters and also remove single character words from the text segments.

\textit{Finally}, we extract vectors for the segment text by taking the union of TF-IDF~\cite{tf-idf} vector and a vector of hand-crafted features obtained from \cite{story2019AAAI}. 
We choose TF-IDF because it is the most popular weighting scheme in the domain of text mining and information retrieval~\cite{beel2016paper,chakraborty2014text}. We create the TF-IDF vector using \texttt{TfidfVectorizer}~\cite{tfidf} with English stop words from the Natural Language Toolkit (NLTK) corpus. The vector of hand-crafted features comprises boolean values corresponding to the presence or absence of custom strings in the training dataset. For example, we use the strings \texttt{cookie, web beacon,} and \texttt{tracking pixel} to indicate the disclosure of \texttt{Identifier\_Cookie} practice.

\vspace{-0.2cm}
\subsubsection{Training Data:} To build a training dataset, we leverage the work of Zimmeck et al. ~\cite{zimmeckEtAlMAPS2019} where they created a privacy policy corpus of 350 most popular mobile apps (APP-350 corpus~\cite{APP350}). 
All the 350 apps selected have more than 5 million downloads. The privacy policies were annotated by legal experts to identify the privacy practices mentioned in the policy text. The annotated label on a privacy policy comprises three parts (or tiers): \textit{(i)} a general or specific practice (twenty-eight unique practices in total), \textit{(ii)} whether that practice has been performed or not, and \textit{(iii)} whether that practice is associated with first or third-party. In first-party practice, the PII is accessed by the code of the app itself, and in third-party practice, the PII is accessed by third-party libraries (such as analytics, advertisements, or social networks). The annotation labels are assigned to the segments (or paragraphs) rather than the whole policy text. The union of labels for the segments provides the disclosure in the whole policy text. We randomly split the APP-350 corpus into training data (n=250) and test data (n=100). 

\vspace{-0.2cm}
\subsubsection{Classification Problem:} For a given privacy policy, we aim to detect the disclosure of privacy practices. A privacy practice comprises three components: \textit{(i)} type of PII (e.g., \texttt{Contact}), \textit{(ii)} negation or approval of the practice being performed (\texttt{Performed} or \texttt{Not Performed}), and \textit{(iii)} party (\texttt{1stParty} or \texttt{3rdParty}). 
We characterize this task as a multi-label text classification problem.

We sub-divide the classification task to identify: \textit{(i)} PII type, \textit{(ii)} procedure i.e., practice has been performed or not, and \textit{(iii)} first-party or third-party. If we make classifier for each unique combination, we will require 112
classifiers for each combination of PII type, perform/not perform, and first/third-party. The limitation of this approach is that the training samples for most of these combinations are limited (less than 100). Therefore, after sub-dividing the classification problem, we only require thirty-two classifiers (twenty-eight for unique PII types, two for procedure (\texttt{Performed} or \texttt{Not Performed}), and two for \textit{parties}). For example, \texttt{Contact\_Email\_Address Performed 3rdParty} will be assigned to a text segment for which \texttt{Contact\_Email\_Address, Performed,} and \texttt{3rdParty} classifiers all return a positive value for that text segment. We consider a policy text discloses a privacy practice if at least one segment in the text returns positive values for at least one PII type classifier, \texttt{Performed} classifier, and \texttt{1stParty} and/or \texttt{3rdParty} classifiers.

We utilize One-vs-the-rest (OvR)~\cite{oneVsRest} classification strategy and test our approach with the Multinomial Naive Bayes~\cite{naiveBayes}, Logistic Regression~\cite{logReg}, and Linear Support Vector Classifiers (SVC)~\cite{SVC}. We select Linear SVC as our classifier for unseen policies' text due to its superior performance on the test dataset (n=100). In particular, the average F1 score (\%) for all classifiers with Multinomial Naive Bayes and Logistic Regression is 17.37\% and 67.54\%, respectively, while for Linear SVC we obtain an average F1 score of 73.75\%. Table~\ref{tab:classifiers}  
shows the performance of our classifiers. Our approach achieves high accuracy, ranging from 91.16\% to 100\% for all the classifiers. 

\vspace{-0.3cm}
\subsection{Static Analysis}
\label{subsec:static}
Resources of an app that are required to run it on a device are bundled together in an Android Package Kit (APK). In the static analysis of mobile apps, after downloading the APK, we decompile it using Apktool~\cite{apktool}. Decompiling APK yields us the assets (including byte-code in the DEX format) and metadata (in the XML format). After disassembling .dex files into smali format, we extract the API calls. We also extract the app permissions from the \texttt{AndroidManifest.xml} file. If an app needs access to a resource outside its sandbox, then it will request the appropriate permission in the \texttt{AndroidManifest.xml} file. 
API calls that do not have the required permissions are not executed. For example, the API \texttt{android.telephony.TelephonyManager.getImei} requires READ\_PRIVILEGED\_PHONE\_STATE permission to retrieve the IMEI (International Mobile Equipment Identity). In particular, we utilize the Android APIs from ~\cite{zimmeckEtAlMAPS2019} and obtain their required permissions from Android API reference~\cite{refAPI}. If we observe an API call for a particular privacy practice in the source code, we check if the required permissions are also requested in the app manifest file. For a given privacy practice, we identify that practice is being performed if both the relevant API call(s) and the required permission(s) exist. 

We distinguish the API \textit{function} calls into system calls, first-party calls, and third-party calls. Calls made by Android class are classified as system calls. We differentiate between first and third-party calls by comparing the classes based on Java's reversed internet domain naming convention~\cite{javaNaming}. The \texttt{.smali} file's package name is matched with the app's package name. If \textit{(i)} both top and second level domains match , or \textit{(ii)} the \texttt{.smali} file's package looks obfuscated (e.g., \texttt{b/a/y.smali}), we classify the API call as first-party. Otherwise, it is classified as a third-party call. In our analysis, we only consider the API calls made by first or third-parties.

Besides obtaining the third-party ``domains'' invoking the API calls, we need to extract the corresponding company names in order to check if these appear in the privacy policy text or not.
Since a company can acquire multiple domains, we leverage the previous work of Binns et al.~\cite{binns2018ownership} in the domain-company ownership. For example, if the API call is from \texttt{adsense.com} then we check for the existence of terms ``adsense'', ``google'' and ``alphabet'' in the privacy policy text. If we find any of the terms in the policy text, we consider the third-party domain as disclosed.

\vspace{-0.3cm}
\subsection{Dynamic Analysis}
\label{subsec:dynamic}

A limitation of static analysis is its inability to capture dynamically loaded code and analyze obfuscated function calls~\cite{lindorfer2014andrubis}. To overcome this limitation, following the work by Ren et al.~\cite{ren2018appversions}, we complement our approach with the dynamic analysis of apps. We employ ReCon~\cite{recon_tool}, a transparency control tool that relies on the machine-learning classification to identify leaks of privacy-related information in the mobile-app traffic~\cite{ren2016recon}. In particular, using a public dataset of annotated traffic flows~\cite{recon_data}, we train a machine-learning classifier (C4.5 decision tree) to predict whether traffic flow is leaking PII. As a feature set, the classifier takes a concatenation of the text of URI, Referrer, postData text, and all other HTTP headers in the flow. The validation accuracy of the classifier is 97.2\% (AUC=0.987), with 97.8\% precision and 96.6\% recall. For those flows predicted to leak (any) PII, we extract the performed privacy practices by matching the feature set against a predefined set of keywords and regular expressions from ReCon~\cite{recon_regex}. 

The gold standard for identifying privacy leaks is by manually logging into the apps and interacting with them. However, this approach is impractical at scale. To automate the analysis, we rely on Android’s UI/Application Exerciser
{\tt Monkey}~\cite{monkey}, a command-line tool that generates pseudo-random user events such as swipes, clicks, or touches. While running an app, we use {\tt mitmproxy}~\cite{mitmproxy} (a TLS-capable interception proxy) to capture all app-generated traffic (HTTP and HTTPS) on a dual-stack~\cite{dual-stack} WiFi in our testbed. Prior work showed that synthetic usage patterns could lead to underestimating the number of privacy leaks compared to manual (human) interactions~\cite{ren2016recon}, since random streams of Monkey events may miss some function calls. While this is a common drawback of automation approaches, Android's Monkey exhibits the best code coverage among existing automation tools~\cite{Zheng2017}.
\begin{table}[t!]
\small
\caption{\small Label mapping between PII leaks from static analysis (\textit{static} leaks) and leaks from dynamic analysis (\textit{dynamic} leaks). 
} 
\centering
\resizebox{.7\columnwidth}{!}{%
\begin{tabular}{|l|l|}
\hline
\textbf{Static Leaks} & \textbf{Dynamic Leaks} \\ \hline
Contact & firstName, lastName \\ \hline
Contact E Mail Address & email (+ hash) \\ \hline
Contact Password & password (+ hash) \\ \hline
Contact Phone Number & phone number \\ \hline
Demographics Gender & gender \\ \hline
Identifier & hardware serial (+ hash) \\ \hline
Identifier \{Ad ID, Cookie\} & \begin{tabular}[c]{@{}l@{}}gsf id, advertiser id (+ hash)\end{tabular} \\ \hline
Identifier Device ID & android id (+ hash)\\ \hline
Identifier IMEI & imei (+ hash) \\ \hline
\begin{tabular}[c]{@{}l@{}}Identifier \{IMSI, Mobile Carrier,\\ SIM Serial\}\end{tabular} & sim id \\ \hline
Identifier MAC & mac addr (+ hash) \\ \hline
\begin{tabular}[c]{@{}l@{}}Location \{Bluetooth, Cell Tower, \\ GPS, IP Address, WiFi\}\end{tabular} & location \\ \hline
\end{tabular}}
\label{tab:mapping}
\vspace{-0.4cm}
\end{table}

In our effort to capture most---if not all---the privacy leaks produced by an app, we take the union of leaks from static and dynamic analysis. Since labels are different in the two cases, we map dynamic leak labels onto the static ones. To this end, we compare the descriptions of static leak labels (\textit{see} Table~\ref{tab:PII_desc}) with the ones of dynamic leak labels in~\cite{ren2018appversions}. Table~\ref{tab:mapping} lists the conversion between the leak labels. We keep the static leak labels as our final ones since they match the labels in the privacy policy text, making them convenient for the following subsection. 

\vspace{-0.3cm}
\subsection{Compliance Analysis}
\label{subsec:compliance}
In the compliance analysis, we compare the leak labels with the privacy policy labels. We consider a compliance violation to have occurred if there is a positive value for a leak label and non-positive value(s) for the corresponding policy text label(s). For some dynamic leaks labels in Table \ref{tab:mapping}, a positive value for a leak label is mapped to multiple labels in static leaks. For example, a positive value for \texttt{gsf id} will result in mapping of positive values for both \texttt{Identifier Ad ID} and \texttt{Identifier Cookie}. Therefore, if we compare individual combined leak labels only, then we may come across unintended violations. For example, if there are positive values for \texttt{Identifier Ad ID} and \texttt{Identifier Cookie} in the leak labels, but only a positive value for \texttt{Identifier Cookie} in the policy text labels, then our system would be flagging \texttt{Identifier Ad ID} as a violation. Thus, to avoid these unintended violations caused by our mapping table, we consider a  violation to have occurred if the policy labels do not return a positive value for any of the similar labels (both \texttt{Identifier Ad ID} and \texttt{Identifier Cookie} in this example). 

For APKs that leak PII to third-parties, we also check for the existence of those third-party domains in the APK's privacy policy text. We examine the domains that are not frequently mentioned in the privacy policies. Note that the mention of the domain in privacy policy does not necessarily imply that the policy text is indicating the sharing of PII with that domain, but the absence of the domain from the policy text definitely implies that the company or domain of the organization with which the app shares PII is not disclosed. Furthermore, by distinguishing apps according to their categories, we identify the categories that are most transparent (complying with the privacy policies) and categories that have most compliance issues.  

\vspace{-0.2cm}
\section{Analysis and Results}
\label{sec:analysis}
In this section, we discuss our analysis of apps' privacy policies and then present leakages observed via our methodology. Finally, we present an analysis of apps compliance with their privacy policies. 

\vspace{-0.3cm}
\subsection{Privacy Policy Analysis}
\label{subsec:polAnalysis}
\vspace{-0.2cm}
We obtain 3,151 unique privacy policies (P.P.) from the Internet Archive and map them to 7,998 different app versions of 405 apps. We term a privacy policy of an app as unique if it is obtained on a different date for an app. For example, if we collect two privacy policies of an app from Internet Archive on two separate dates, then we term those two P.P. as unique, even if the contents of both the policies are identical. 

To map the P.P. to the different app versions, we bind each APK to the app's privacy policy that is immediately after the APK release date.  
We note that 42.6\% of the analyzed APKs have a time difference of fewer than 12 months between the APK release date and the P.P. date
.

The collected privacy policies contain 143,783 policy segments (or paragraphs) in total. We restrict this set by considering only those segments that return
positive values on \textit{(i)} at least one of the twenty-eight practice classifiers (e.g., Contact), \textit{(ii)} procedure (i.e., the practice described in the segment is being performed or not performed), and \textit{(iii)} parties (i.e., 1st Party or 3rd party). Out of a total of 143,783 policy segments, 19,371 segments return a positive value for the above three categories. We term these segments as {\it valid segments}. The valid segments span across 2,455 privacy policies mapped to 329 apps. We will rely on this subset (valid segments) in the remainder of the analysis to measure privacy policy violations.

\begin{figure}
    \centering
    \subfloat[\label{a}]{\includegraphics[scale=0.38, keepaspectratio]{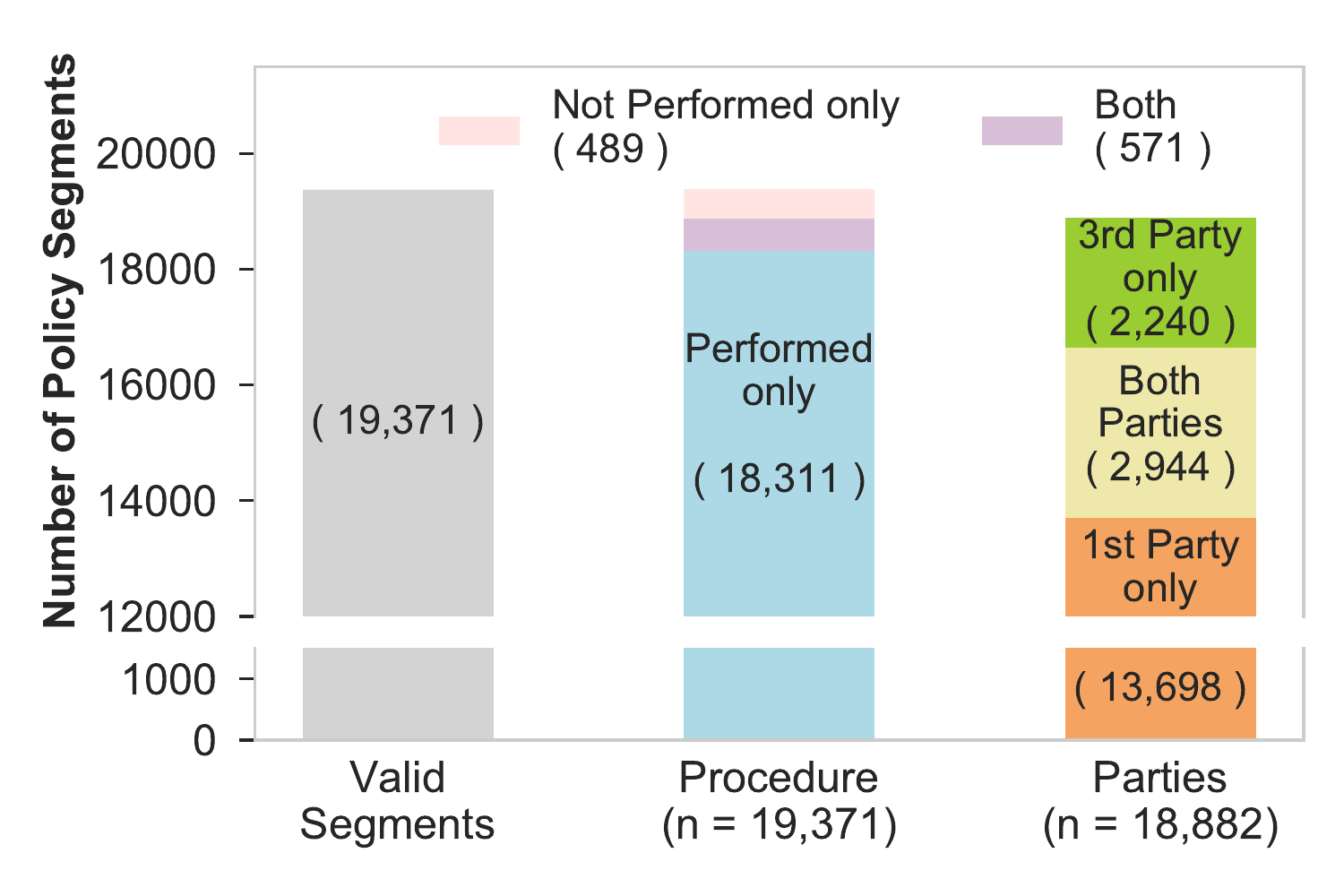}\label{fig:breakdown}}\quad
    \subfloat[\label{b}]{\includegraphics[scale=0.39, keepaspectratio]{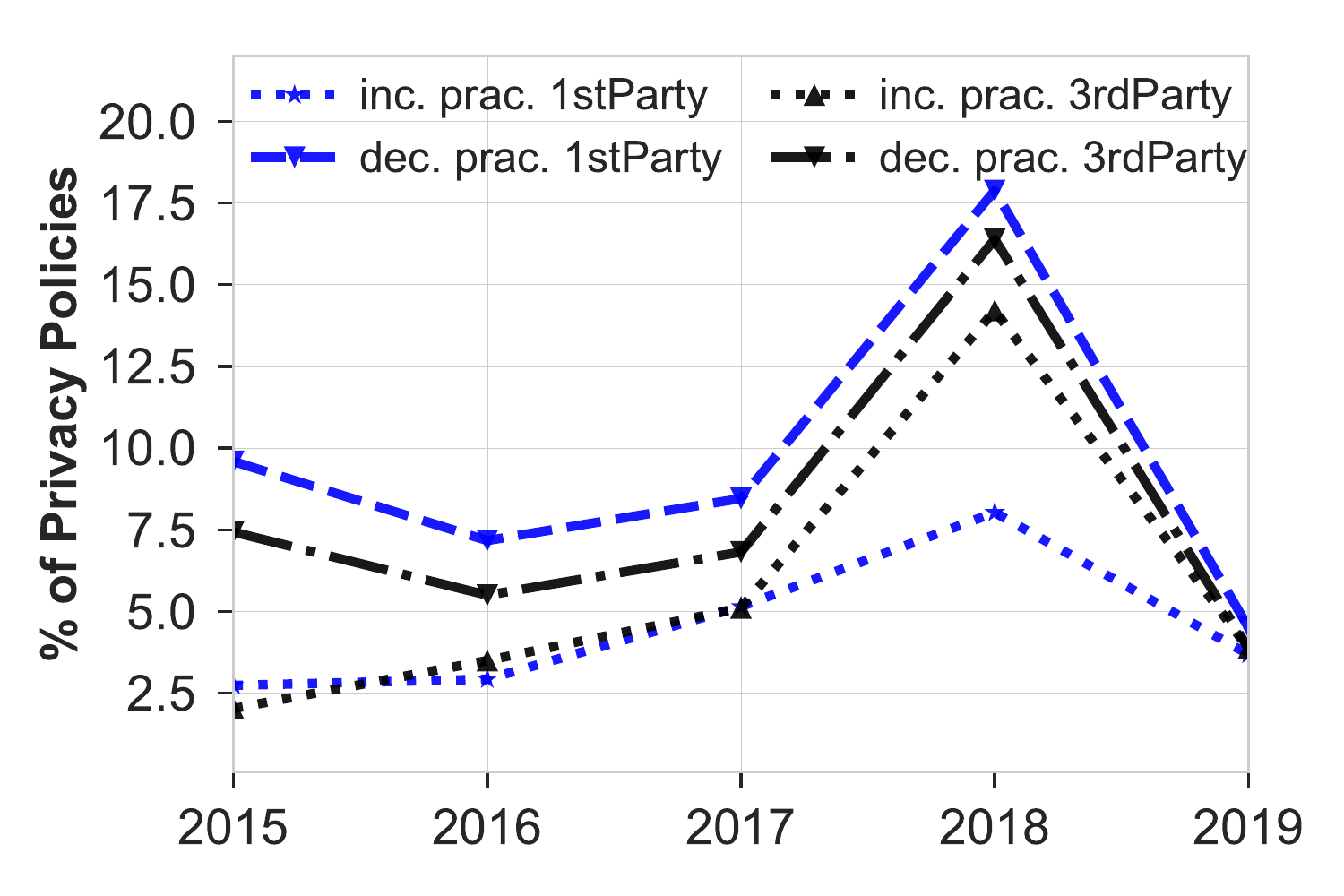}\label{fig:practices_change}}
    \vspace{-0.2cm}
    \caption{\small (a) Breakdown of valid segments (n=19,371) into Procedure (Performed / Not Performed). Valid segments for which Performed classifier returns a positive value (n=18,882; Performed + Both) are broken down further into parties (1st / 3rd). (b) Percent of P.P. versions in the given year showing an increase/decrease in the number of practices disclosed. Among the newer versions of P.P. each year, more number of them continue to disclose less PII.}
    \vspace{-0.3cm}
\end{figure}

Figure \ref{fig:breakdown} shows the distribution of the valid policy segments by the procedure. A vast majority (97.48\%) of valid policy segments have a positive value for the \texttt{Performed} classifier. These segments (n=18,882) span across 2,422 P.P. mapped to 327 apps. This also includes the segments that return positive values for both \texttt{Performed} and \texttt{Not Performed} classifiers. A segment is not contradictory if it has positive values for both {\tt Performed} and {\tt Not Performed}. For example, the segment {\it ``we will only store your email and will not collect your location''} can have positive values returned by both the classifiers. In such cases however, our classifiers are unable to distinguish between {\tt Performed} PII type(s) and {\tt Not Performed} type(s). Therefore, to overcome this limitation and prevent falsely flagging a policy for a violation, we assign all flagged PII types as {\tt Performed}. Similarly, for the segments that return positive values for both the parties (n=2,944), we consider the flagged PII types(s) to be {\tt Performed} by both the parties. Among the segments that return positive value for the {\tt Performed} classifier (n=18,882), most have at least one practice being performed by the first-party (88.14\%).   

The segments that return a positive value for the \texttt{Not Performed} classifier only (n=489) span across the P.P. of 89 apps and 382 APKs. Among them, 33 APKs comprising eight apps have segments that only return a positive value for the \texttt{Not Performed} classifier, i.e., these APKs' P.P. only state that PII-revealing practices are not performed. The three most common practices disclosed in the P.P. are {\tt Identifier Cookie}, {\tt Identifier IP Address}, and {\tt Contact E Mail Address}. 
We empirically observe that the number of practices disclosed in the analysed P.P. is 19,038. This number implies an average of 7.86 per APK, or 7.75 if we also include the APKs whose policy segments only return a positive value for the \texttt{Not Performed} classifier.


For a given year, a P.P. of an app can have multiple versions (based on the date the P.P. was obtained). Among all the P.P. that disclose the collection of PII to 1st party, 86.5\% comprise of multiple versions in the same year (86.1\% for 3rd party, respectively). Among these multiple versions, we compare the P.P. of an app to their preceding version to identify if they are disclosing the collection of more or less PII. Figure~\ref{fig:practices_change} shows the annual disclosure trend of P.P. obtained in the years 2015-2019. We do not consider P.P. before 2015 due to less number of versions available ($<$ 10 annually). We observe the prevalence of P.P. with less PII disclosures compared to their preceding version. In particular, 17.9\% of P.P. in 2018 show decreasing numbers of 3rd party disclosures (respectively, 16.4\% 1st party disclosures), whereas only 8\% of P.P. are found with more 3rd party disclosures (respectively, 14.2\% 1st party disclosures). For apps whose behavior (leaks) remain uniform, the drop in number of disclosures in P.P. can have potential compliance issues. 

\subsection{Analyzing Leakages} 
\label{subsec:leaksAnalysis}
\textbf{Static Analysis:} 
We analyze 7,741 different APKs (or app versions) of 350 different apps, from which we extract the PII-sensitive API calls. Among these calls, 24.2\% (n = 28,001) are made by the first-party whereas 75.8\% calls (n=87,497) are initiated by third-party services or domains. 

\begin{figure}[t]
    \centering
    \subfloat[{}]{\includegraphics[scale=0.39, keepaspectratio]{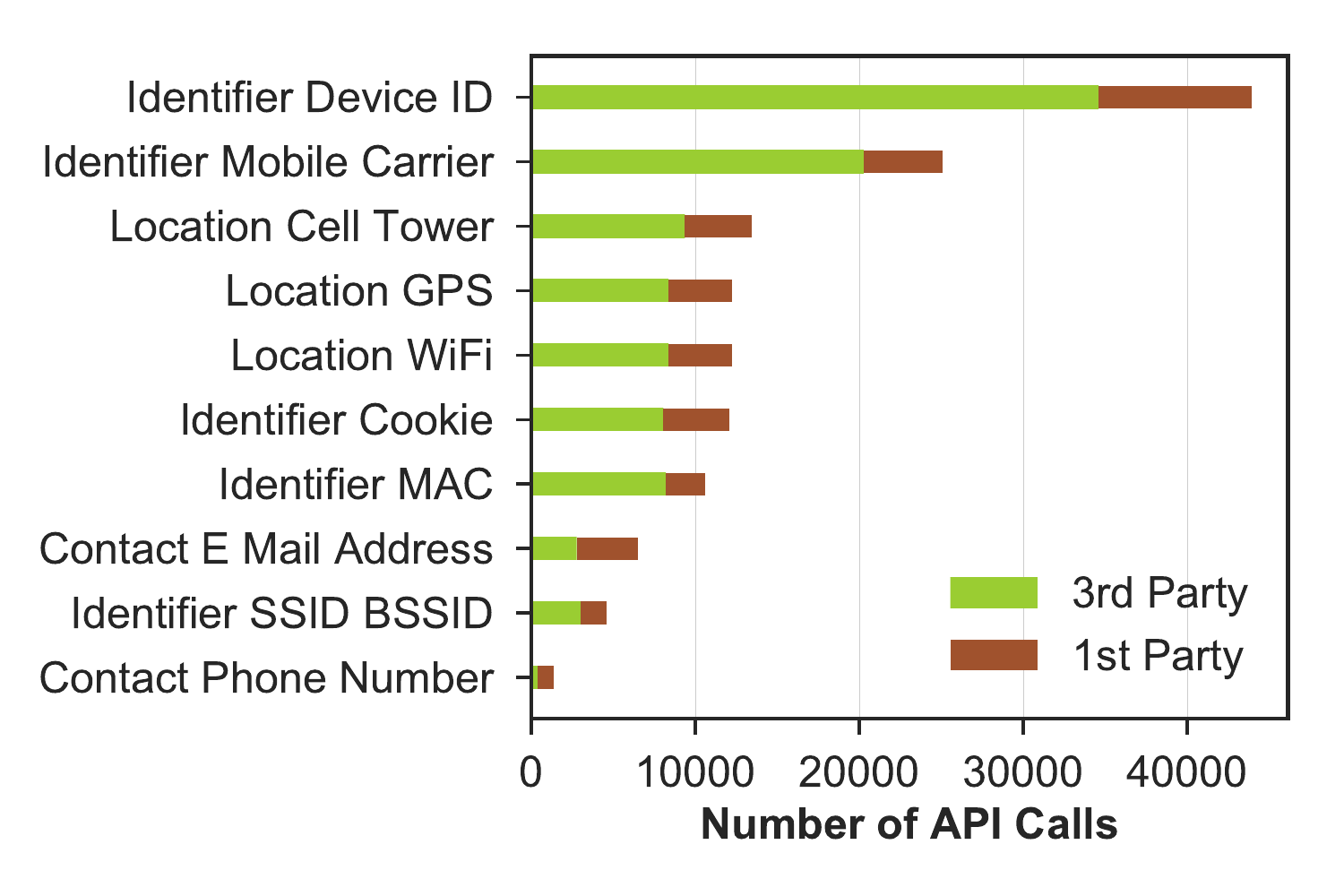}\label{fig:apiCalls}}\quad
    \subfloat[{}]{\includegraphics[scale=0.38, keepaspectratio]{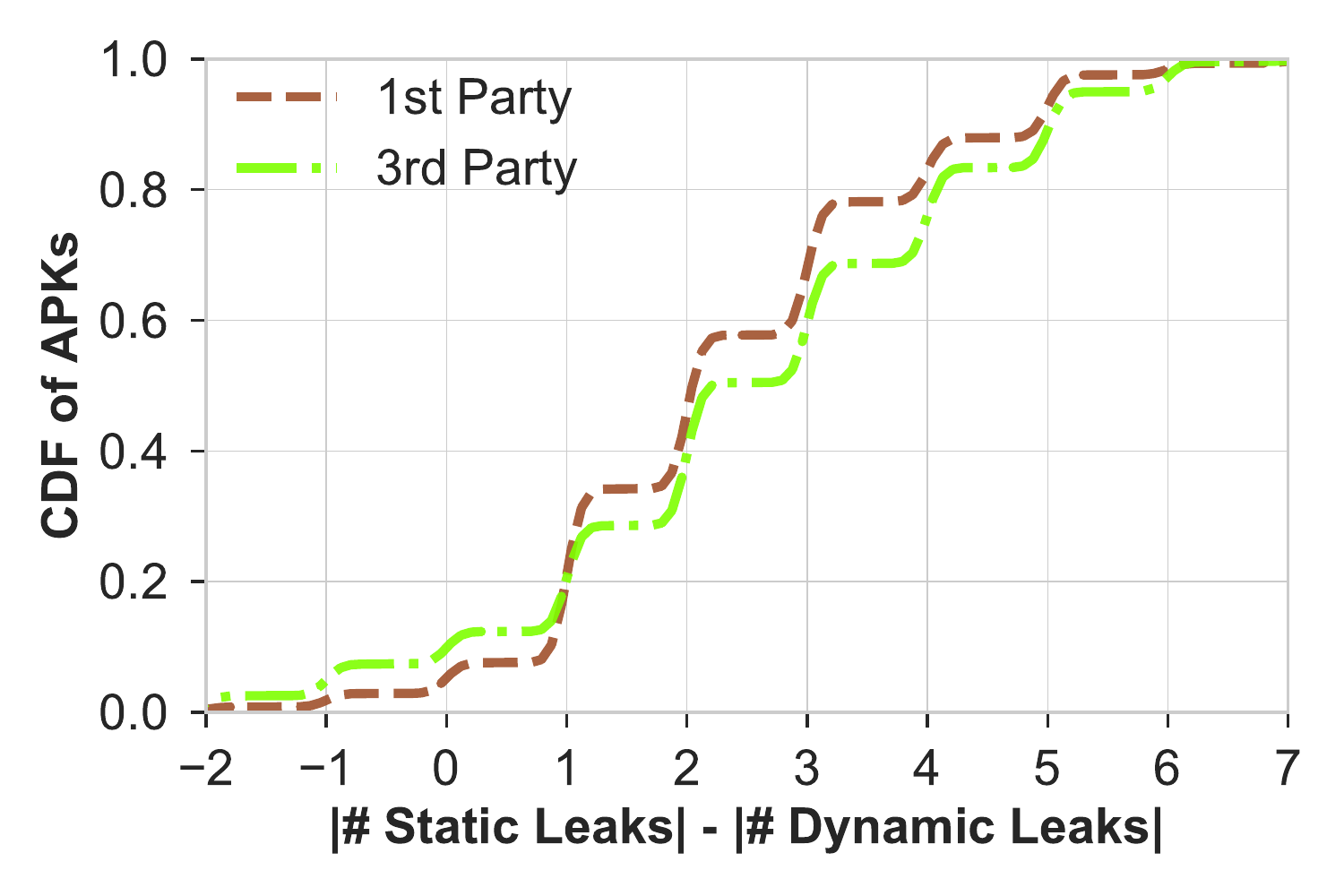}\label{fig:leaks_comparison}}
    \vspace{-0.2cm}
    \caption{\small (a) Number of sensitive API calls observed in the static analysis. (b) Difference in number of static leaks and dynamic leaks. At least 85\% of analyzed APKs have more leaks discovered in the static analysis than in the dynamic analysis.}
\end{figure}

Figure \ref{fig:apiCalls} shows the distribution of these API calls to different types of PII. 
\texttt{Identifier Device ID} and \texttt{Identifier Mobile Carrier} are the top two PII that API calls (from both first and third-parties) are collecting. 

\textbf{Contrasting Static and Dynamic Leaks:}
Table \ref{tab:leaksParties} shows the annual distribution of apps and versions (APKs) that report PII leak(s) through static analysis, dynamic analysis, and both. 

\begin{table}[t]
\caption{{\small Number of unique apps (and versions i.e., APKs) with at least one PII leak. We report the leaks revealed by static and dynamic analysis. We also enumerate the apps (and APKs), which are common to both the analysis. We only consider the apps with retrieved privacy policy.}}
\centering
\resizebox{.7\columnwidth}{!}{%
\begin{tabular}{lrrrrrr}
\toprule
\multirow{2}{*}{\textbf{Year}} & \multicolumn{2}{c}{\textbf{(A) Static }} &
\multicolumn{2}{c}{\textbf{(B) Dynamic}} &
\multicolumn{2}{c}{\textbf{(A $\cap$ B)}} \\
\cmidrule(lr){2-3} \cmidrule(lr){4-5} \cmidrule(lr){6-7}
 & \textbf{\#Apps} & \textbf{\#APKs} & \textbf{\#Apps} & \textbf{\#APKs} & \textbf{\#Apps} & \textbf{\#APKs} \\
\midrule
2008 & 3 & 33 & 5 & 44 & 2 & 30 \\
2009 & 6 & 32 & 6 & 18 & 3 & 4 \\
2010 & 24 & 102 & 17 & 34 & 13 & 20 \\
2011 & 47 & 253 & 39 & 119 & 26 & 69 \\
2012 & 87 & 550 & 66 & 197 & 53 & 171 \\
2013 & 136 & 868 & 81 & 334 & 66 & 299 \\
2014 & 173 & 1302 & 110 & 525 & 95 & 476 \\
2015 & 272 & 2035 & 208 & 1018 & 178 & 847 \\
2016 & 301 & 2485 & 235 & 1492 & 197 & 1242 \\
\bottomrule
\end{tabular}}
\label{tab:leaksParties}
\vspace{-0.4cm}
\end{table}

\begin{table}[!ht]
\centering
\small
\caption{{\small Summary of PII Leaks from analyzed apps. In columns 2, 3, and 6, the numbers in parenthesis i.e., () represent the number of versions of apps (APKs).}}
\tabcolsep=0.05cm
\resizebox{0.85\columnwidth}{!}{%
\begin{tabular}{l|l|l|r|r|l|r|r}
\toprule
\multirow{3}{*}{\textbf{Year}} & 
{\bf \#Apps} & 
\multicolumn{3}{c|}{\textbf{Leaks to 1st Party}} & 
\multicolumn{3}{c}{\textbf{Leaks to 3rd Party}} \\
\cmidrule(lr){3-5} \cmidrule(lr){6-8}
 & & \multicolumn{1}{c|}{\textbf{\# Apps}} & \multicolumn{1}{c|}{\textbf{\#Leaks}} & \multicolumn{1}{c|}{\textbf{\#Leaks}} & \multicolumn{1}{c|}{\textbf{\#Apps}} & \multicolumn{1}{c|}{\textbf{\#Leaks}} & \multicolumn{1}{c}{\textbf{\#Leaks}} \\ 
  & \multicolumn{1}{c|}{\bf (\#APKs)} & \multicolumn{1}{c|}{(\#APKs)} & & \multicolumn{1}{c|}{per APK} & \multicolumn{1}{c|}{(\#APKs)} & & \multicolumn{1}{c}{per APK} \\
\midrule
2008 & 6 (47) & 4 (34) & 116 & 3.41 & 5 (34) & 0 & 0 \\
2009 & 9 (46) & 6 (29) & 116 & 4.00 & 6 (21) & 12 & 0.57 \\
2010 & 28 (116) & 20 (84) & 252 & 3.00 & 21 (63) & 187 & 2.97 \\
2011 & 59 (303) & 39 (169) & 558 & 3.30 & 53 (269) & 726 & 2.70 \\
2012 & 99 (576) & 67 (354) & 1127 & 3.18 & 94 (525) & 1531 & 2.92 \\
2013 & 149 (903) & 99 (637) & 2217 & 3.48 & 141 (835) & 3202 & 3.83 \\
2014 & 188 (1351) & 132 (940) & 3289 & 3.50 & 182 (1306) & 5780 & 4.43 \\
2015 & 302 (2206) & 207 (1485) & 5475 & 3.69 & 300 (2155) & 9228 & 4.28 \\
2016 & 339 (2735) & 234 (1841) & 7184 & 3.90 & 336 (2707) & 11998 & 4.43 \\
\bottomrule
\end{tabular}%
}
\label{tab:leaksSum}
\vspace{-0.3cm}
\end{table}

\begin{figure*}[!bht]
\centering
\subfloat[{Leaks to 3rd Party}]{\includegraphics[scale=0.28, keepaspectratio]{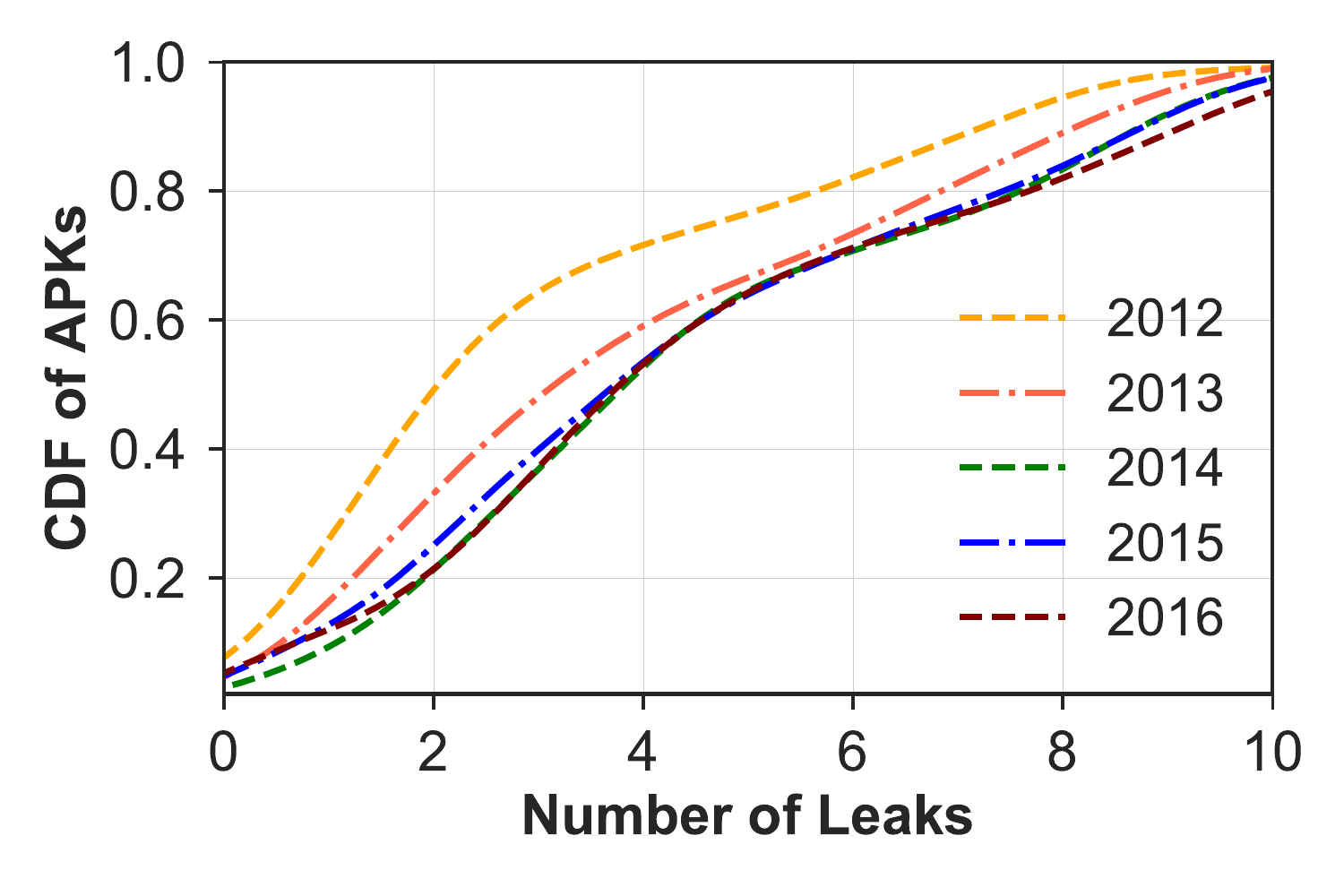}\label{fig:leaks_3rd}}
\subfloat[{Violations to 1st Party}]{\includegraphics[scale=0.28, keepaspectratio]{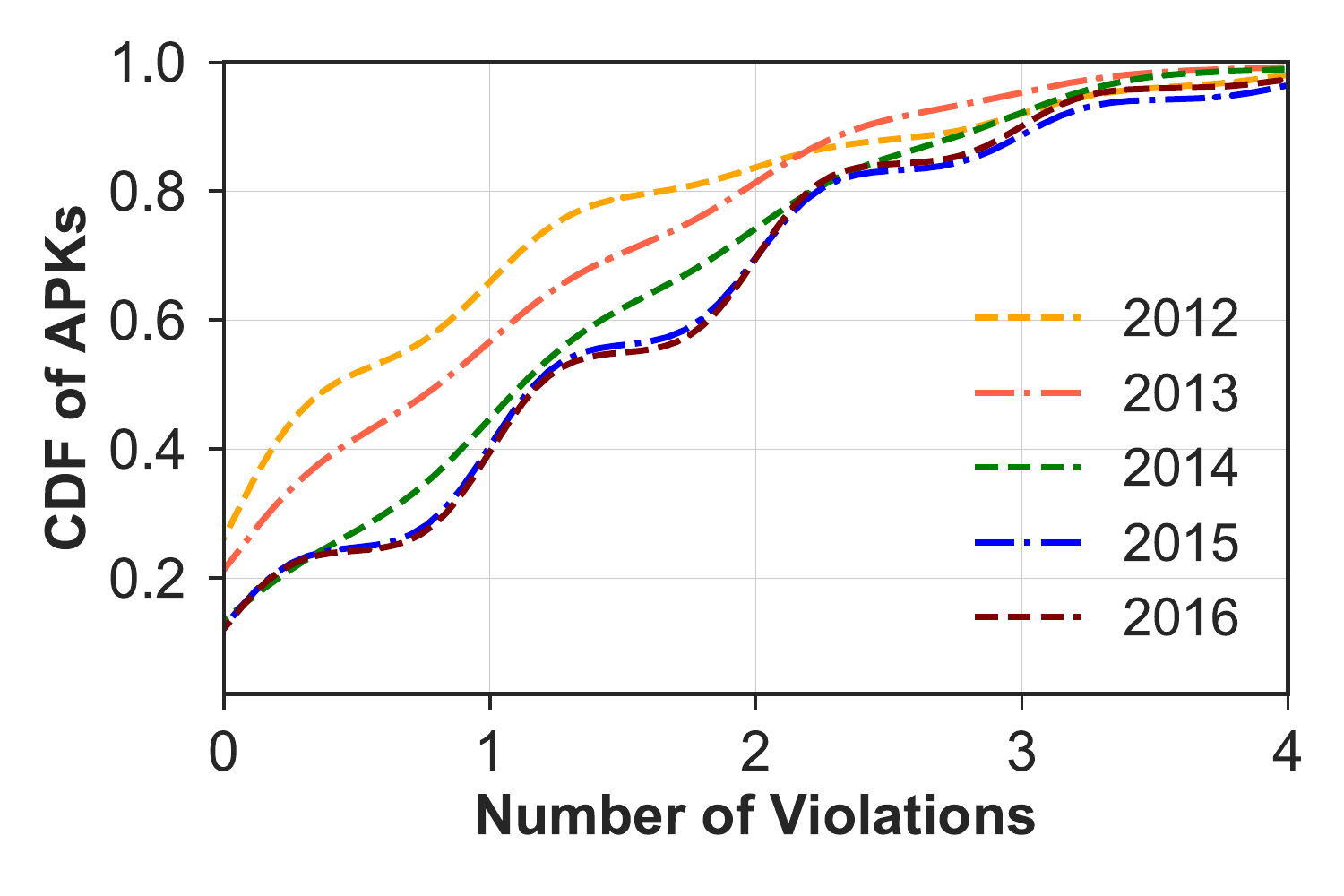}\label{fig:viol_1st}}
\subfloat[{Violations to 3rd Party}]{\includegraphics[scale=0.28, keepaspectratio]{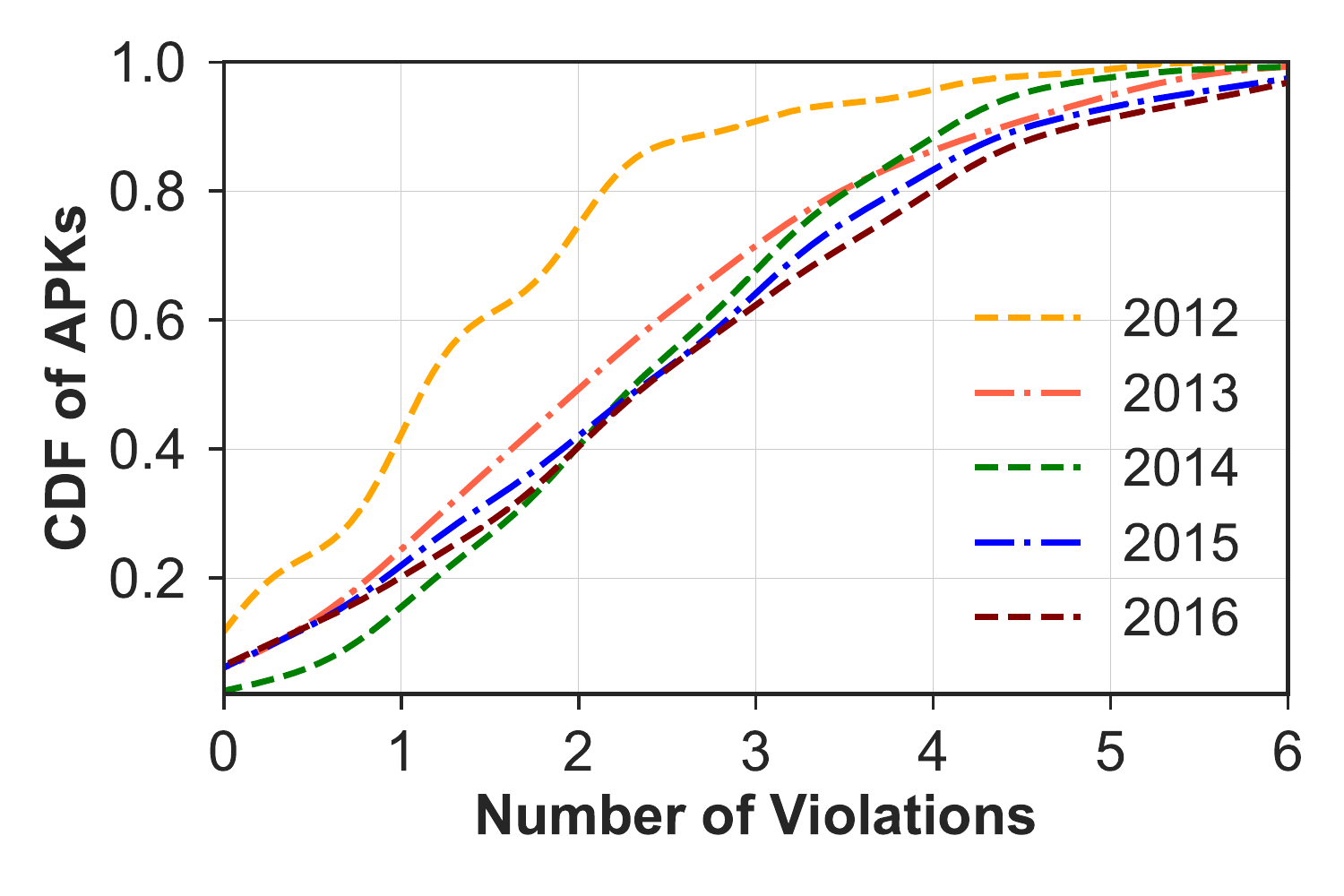}\label{fig:viol_3rd}}
\vspace{-0.2cm}
\caption{{\small Cumulative Distribution Functions (CDFs) of leaks for (a) 3rd party. Annual CDF of violations for (b) 1st party, and (c) 3rd party. Over time, the total number of leaks from APKs and PII violations are increasing.}}
    \label{fig:leaks_violations}
\vspace{-0.1cm}
\end{figure*}

We observe significant differences in the sets of PII-leaks found by static analysis and dynamic analysis, with the former revealing more leaks than the latter in the majority of cases ($>$85\% of analyzed APKs) as shown in Figure \ref{fig:leaks_comparison}. 
For example, \texttt{com.ace.cleaner} version 6.0 leaked \texttt{Identifier Device ID, Contact E Mail Address, Identifier MAC,} and \texttt{ Identifier Mobile Carrier} to third-parties in the static analysis whereas only \texttt{advertiser id,} and \texttt{location} were leaked to third-parties in the dynamic analysis. 
Part of the PII leaks is missed by dynamic analysis due to limitations of Android’s UI/Application Exerciser Monkey in triggering all PII-related API calls. At the same time, we observe that static analysis could not detect some PII leaks (user's ad ID, cookie ID, location data sharing) as these appeared in dynamically loaded code or obfuscated function calls. 
The limitations of both static and dynamic analysis demonstrate that only the union set of leaks from the static and dynamic analysis can allow for effective or useful profiling of the privacy-related behavior of an app.

Besides, we examine the prevalence of detection from either static or dynamic analysis for different PII leaks in the period 2012-2016 -- we do not analyze the years before 2012 due to insufficient data ($<$350 APKs). 
We empirically observe that the most commonly accessed PII is the \texttt{Identifier Device ID} by first-parties and \texttt{Identifier Mobile Carrier} and \texttt{Identifier Device ID} by third-parties.

{\bf Trend of Leaks:}
We then take the union of PII leaks observed in static and dynamic leaks using the mapping table (\textit{see} Table \ref{tab:mapping}). The summary of combined leaks, shown in Table \ref{tab:leaksSum}, suggest that the average number of PII leaks per APK has risen consistently since 2012, from an average of 3.18 to first-parties (respectively, 2.92 to third-parties) in 2012 to 3.9 (respectively, 4.43 to third-parties) in 2016. Another interesting takeaway is the rise of leaks to third-parties in comparison to the first-parties. Till 2012, apps leaked PII mostly to the first-parties, and since then, they have been leaking mostly to third-parties. We can also observe this trend in Figure~\ref{fig:leaks_violations}, which shows a more significant increase over time for the third-party leaks (Figure~\ref{fig:leaks_3rd}) compared to the case of first-party leaks
.  
For instance, 62.9\% of the APKs exhibit less than three leaks to third-parties in 2012, compared to 27.3\% in 2016. Similarly, 21.7\% of the APKs exhibit more than five leaks to third-parties in 2012 which increased to 31.1\% in 2016.
This trend can be attributed to the increasing embedding of ad and tracking libraries in the apps~\cite{AndroidLeaks,viennot2014measurement}.

\begin{table}[!t]
\vspace{-0.3cm}
\tabcolsep=0.05cm
\footnotesize
\caption{{\small Summary of contradictions or violations of analyzed apps with their privacy policies. In column 2, the numbers in parenthesis i.e., () represent the number of app versions (APKs).
}}
\centering
\resizebox{.95\columnwidth}{!}{%
\begin{tabular}{r|l|l|r|r|r|r|r}
\toprule
\multirow{4}{*}{\textbf{Year}} &
\multicolumn{1}{c|}{\textbf{\#Apps}} & \multicolumn{1}{c|}{\textbf{\#(\%)APKs}} & \multicolumn{1}{c|}{\textbf{\#Leaks}} & \multicolumn{1}{c|}{\textbf{\#Leaks}} & \multicolumn{1}{c|}{\textbf{\#Leaks}} & \multicolumn{1}{c|}{\textbf{\#Leaks}} & \multicolumn{1}{c}{\textbf{\#Leaks}} \\
 & \multicolumn{1}{c|}{\textbf{(\#APKs)}} & \multicolumn{1}{c|}{\textbf{comply.}} & \multicolumn{1}{c|}{\textbf{Total}} & \multicolumn{1}{c|}{\textbf{viol.}} & \multicolumn{1}{c|}{\textbf{per APK}} & \multicolumn{1}{c|}{\textbf{to 1st P.}} & \multicolumn{1}{c}{\textbf{to 3rd P.}} \\
  & \multicolumn{1}{c|}{\textbf{with $\geq$ 1}} & \multicolumn{1}{c|}{\textbf{with P.P.}} & & \multicolumn{1}{c|}{\textbf{P.P.}} & \multicolumn{1}{c|}{\textbf{viol. P.P.}} & \multicolumn{1}{c|}{\textbf{viol. P.P.}} & \multicolumn{1}{c}{\textbf{viol. P.P.}}\\
   & \multicolumn{1}{c|}{\textbf{valid seg.}} & \multicolumn{1}{c|}{\bf} &  & \multicolumn{1}{c|}{\bf} & \multicolumn{1}{c|}{\bf } & \multicolumn{1}{c|}{\bf } & \multicolumn{1}{c}{\bf }\\ 
\midrule
2008 & 3 (11) & 11 (100) & 8 & 0 & 0 & 0 & 0 \\
2009 & 5 (24) & 18 (75) & 57 & 9 & 0.38 & 0.38 & 0\\
2010 & 18 (80) & 37 (46.25) & 313 & 71 & 0.89 & 0.56 & 0.32 \\
2011 & 37 (187) & 62 (33.16) & 820 & 240 & 1.28 & 0.70 & 0.59 \\
2012 & 69 (364) & 117 (32.14) & 1622 & 556 & 1.53 & 0.65 & 0.88 \\
2013 & 96 (584) & 104 (17.81) & 3473 & 1264 & 2.16 & 0.77 & 1.39 \\
2014 & 126 (860) & 79 (9.19) & 5332 & 2160 & 2.51 & 1.00 & 1.51 \\
2015 & 195 (1381) & 149 (10.79) & 8640 & 3909 & 2.83 & 1.10 & 1.73 \\
2016 & 225 (1748) & 188 (10.76) & 11338 & 5210 & 2.98 & 1.04 & 1.94 \\
\bottomrule
\end{tabular}}
\label{tab:violSum}
\vspace{-0.3cm}
\end{table}

\vspace{-0.3cm}
\subsection{Compliance Analysis}
\label{subsec:compAnalysis}
\vspace{-0.1cm}
\textbf{Disclosure of practices performed:} In the compliance analysis, we compare the combined (static and dynamic) leaks of an APK with the practices reported in the privacy policy. A privacy policy violation occurs if a leak observed in static/dynamic analysis is not mentioned in the privacy policy. For example, our analysis finds that the app {\tt com.fitbit.FitbitMobile}~\cite{fitbit} with the version code 2182996 (APK release date: 22 Jun 2016) has privacy policy (dated: 16 Sep 2016)  violations for \texttt{Identifier Device ID}, and \texttt{Identifier Mobile/Sim} practices being performed by the third-parties. On manually inspecting the privacy policy text, we do not find the disclosure of the above-mentioned practices for third-parties. For disclosing data to third-parties, the app's policy states: {\it ``Data That Could Identify You Personally Identifiable Information (PII) is data that includes a personal identifier like your name, email or address, or data that could reasonably be linked back to you. We will only share this data under extremely limited circumstances''}. This generic clause may seem to be allowing the app developers to share the user's data, but Android's policy on user data states that the developers must describe the data being collected and explain their usage~\cite{googlePlayDisclosure}. This means that the PII sharing from Fitbit app does not comply with the declared public privacy policy of the app. 

\begin{figure}[t]
    \centering
    \includegraphics[scale=0.41, keepaspectratio]{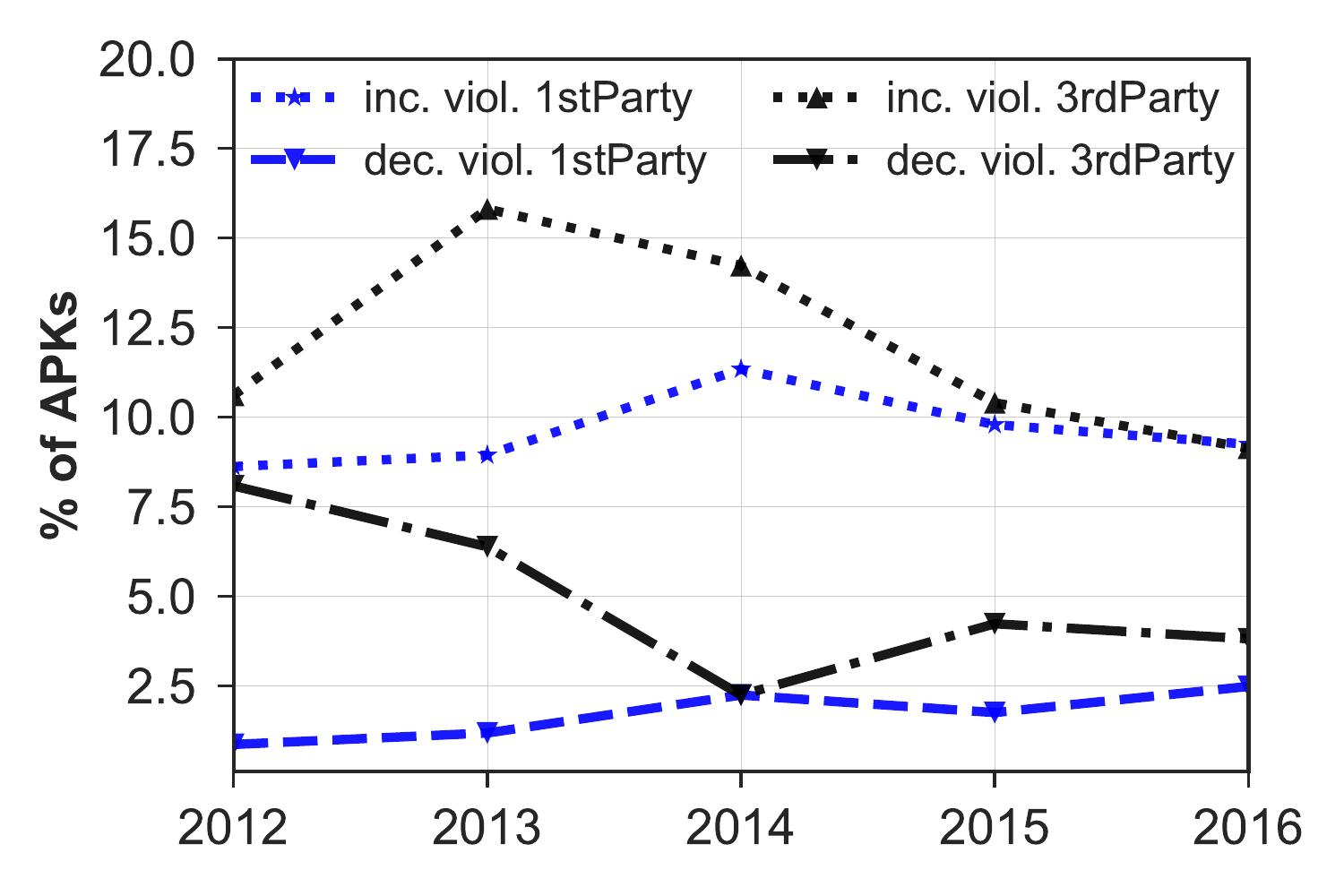}
    \vspace{-0.3cm}
    \caption{{\small Percent of APKs released in the given year showing an increase/decrease in the number of violations compared to their preceding version (APK). Newer versions of apps continue to leak more PII that violate privacy policy.}}
    \label{fig:violChange}
    \vspace{-0.3cm}
\end{figure}

Table \ref{tab:violSum} lists the annual summary of violations observed in APKs. These violations only refer to those APKs whose privacy policy contains at least one valid segment i.e., a segment that returns positive values on \textit{(i)} at least one of the twenty-eight practice classifiers (e.g., Identifier Cookie), \textit{(ii)} procedure (i.e., the practice being described in the segment is being performed or not performed), and \textit{(iii)} parties (i.e., 1st Party or 3rd party). Most of the policy text segments (124,412/143,783) that we classified did not return positive values for the above three categories (\textit{see \S \ref{subsec:polAnalysis}}). 
For these cases, our machine-learning classifier could not determine whether the policy text presents a practice as ``performed" or ``not performed" by the parties. 
From Table \ref{tab:violSum}, we can observe that the compliance of APKs with their P.P. has decreased considerably from 33.16\% in 2011 to 10.76\% in 2016. The average number of leaks per APK that is not disclosed in the P.P. is also steadily increasing. We can observe this trend in Figure \ref{fig:viol_1st} and \ref{fig:viol_3rd}, showing that the number of P.P. violations is consistently increasing over time for a significant fraction of APKs.


We empirically note that \texttt{Identifier Mobile/Sim} (grouping of \texttt{Identifier IMSI, Identifier Mobile Carrier,} and \texttt{Identifier SIM Serial} based on the mappings in Table \ref{tab:mapping}) annually contributes to more than 25\% of first-party and more than 30\% of third-party violations. \texttt{Identifier Device ID} is next, comprising more than 20\% of annual violations. This suggests that privacy policy violations are often due to apps {\it fingerprinting} users' mobile devices without revealing it. 


\textbf{Compliance of an app across newer versions:} For a newer version (APK) of an app released, it can have \textit{(i)} equal number, \textit{(ii)} higher number, or \textit{(iii)} smaller number of privacy policy violations compared to the preceding version of the given app. Figure \ref{fig:violChange} shows the annual compliance trend of APKs released in the years 2012-16 compared to their preceding version. For example, if an app released versions in Oct'15, Dec'15, Feb'16, Jul'16, and Nov'16, then for the year 2016 the number of violations of APK released in Nov'16 will be compared with Jul'16, Jul'16 will be compared with Feb'16, and Feb'16 will be compared with Dec'15. 
For the set of app versions released in the years 2012-2016, we observe the prevalence of apps with more privacy policy violations compared to their preceding version. In particular, 9.1\% of APKs from 2016 show increasing numbers of third-party violations (respectively, 9.2\% first-party violations), whereas only 3.8\% of APKs are found with less third-party violations (respectively, 2.5\% first-party violations).

\begin{table}[]
    \caption{{\small  Breakdown of privacy policies based on the disclosure of 3rd party domains with which their APKs share PII.}}
    \centering
    \vspace{-0.3cm}
    \tabcolsep=0.25cm
    \resizebox{.75\columnwidth}{!}{%
    \begin{tabular}{c|c|c}
    \toprule
        \multicolumn{3}{c}{\textbf{\% (\#) of APKs with Privacy Policies (P.P.) mentioning}} \\
        \cmidrule(lr){1-3}
        \textbf{ALL} & \textbf{NONE} & \textbf{PARTIAL} \\
        \textbf{3rd party domains} & \textbf{3rd party domains} & \textbf{3rd party domains} \\
    \midrule
    17.8\% (1,419) & 23.85\% (1,901) & 58.35\% (4,651) \\
    \bottomrule
    \end{tabular}}
    \label{tab:domainsDisclosure}
\end{table}

\begin{figure}[!b]
\vspace{-0.8cm}
    \centering
    \subfloat[{}]{\includegraphics[scale=0.37, keepaspectratio]{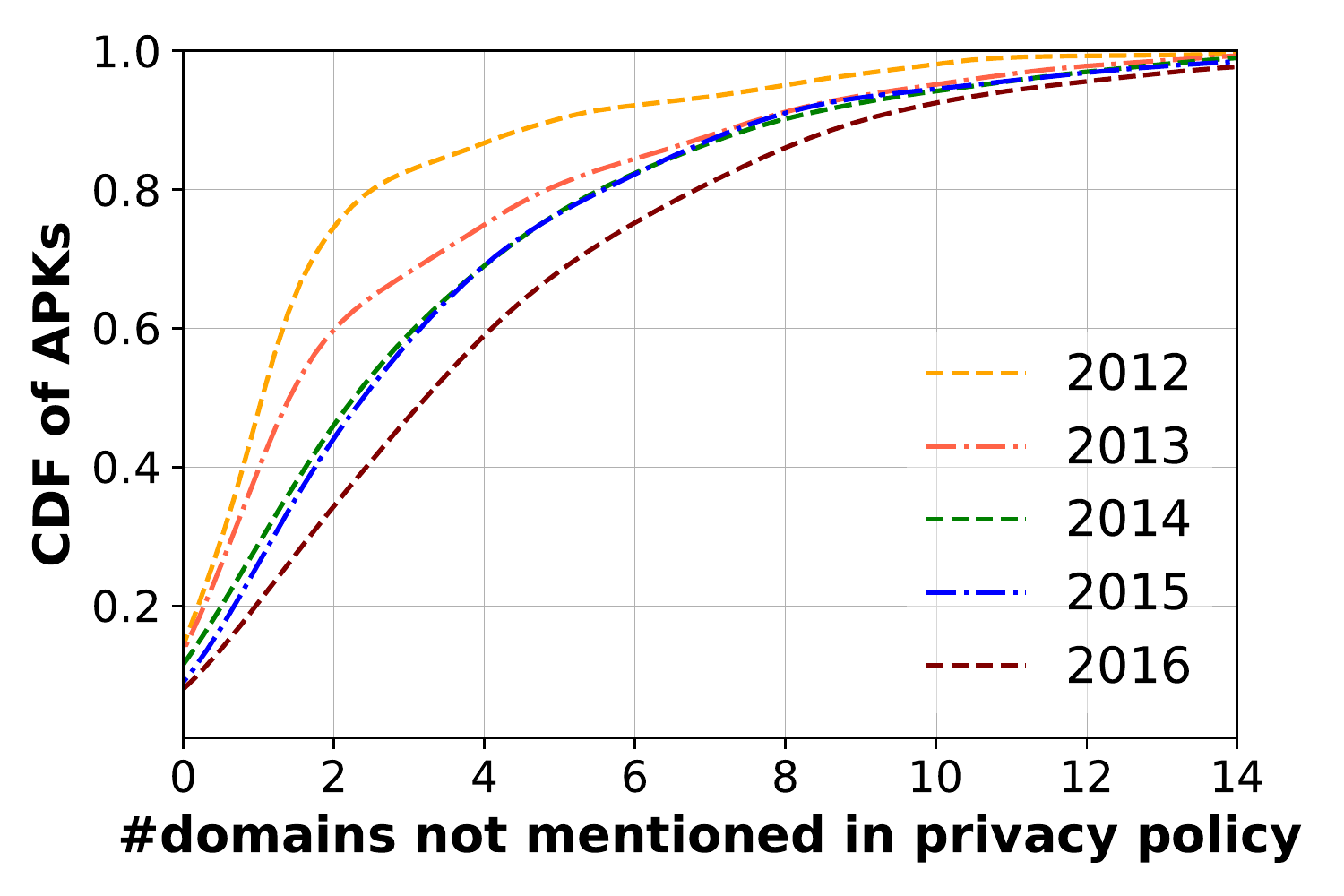}\label{fig:domsNotMent}}
    \subfloat[{}]{\includegraphics[scale=0.38, keepaspectratio]{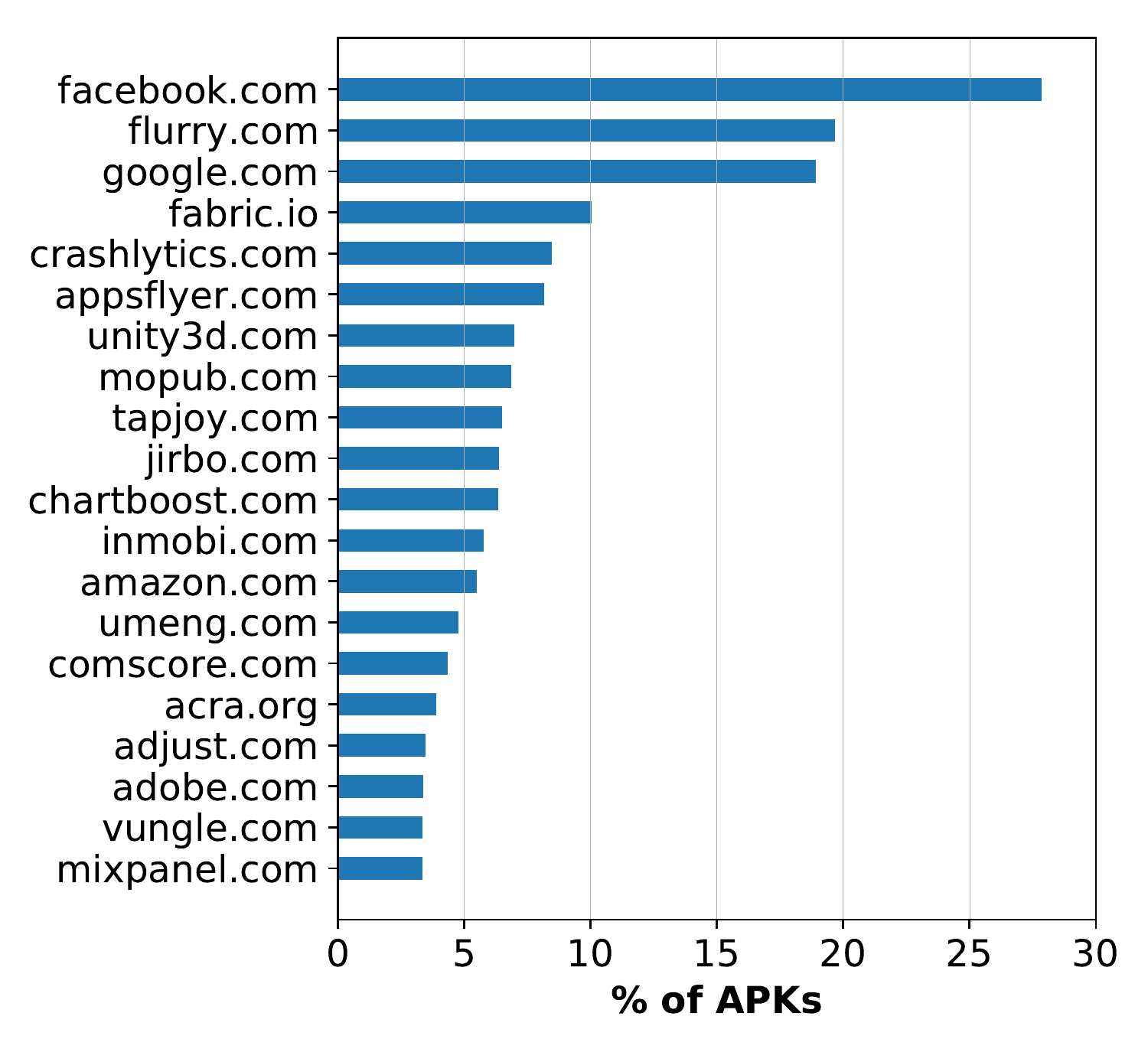}\label{fig:domsDistribution}}
    \vspace{-0.3cm}
    \caption{{\small (a) CDFs of number of domains not mentioned in the privacy policy of APKs. Each subsequent year, APKs share PII to more 3rd party domains that are not disclosed in their privacy policy. (b) Prevalence of undisclosed domains among the privacy policies of APKs. 
    }}
    \vspace{-0.4cm}
\end{figure}

\vspace{-0.2cm}
\textbf{Disclosure of 3rd Party Domains:} Given the set of PII leaks to third-party domains, we verify whether those domains are mentioned in the P.P. text. Table \ref{tab:domainsDisclosure} shows the distribution of P.P. based on the disclosure of third-party domains. We note that a vast majority of P.P. (82.2\%) do not mention at least one domain with which they share PII. 

Figure \ref{fig:domsNotMent} shows the annual distribution of the number of domains that are not mentioned in the P.P. 
What we observe is a significant increase in the number of third-party domains that are not disclosed by app publishers despite being involved in PII sharing. 
In particular, while in 2012, almost 70\% of the released APKs had no more than one 3rd party domain ``missed" by their privacy policies; in 2016 this percentage has dropped to less than 30\%. 
Figure \ref{fig:domsDistribution} shows the twenty most recurrent domains that are not disclosed in the privacy policies. Overall, these twenty domains account for 49.5\% of the total instances where third-party domains are not disclosed. Each of these domains provides a library for analytical services, advertisements, social networking, or utility/developer tools to the app. The most frequently not mentioned domain in the privacy policies by app publishers is \texttt{facebook.com}, accounting for more than a quarter (27.8\%) of the analyzed APKs. It is followed by \texttt{flurry.com} (19.7\%) and \texttt{google.com} (18.9\%).  

\vspace{-0.2cm}
\section{Discussion}
\label{sec:disc}
\vspace{-0.2cm}
The increase in the number of violations over time is concerning. The causes of increase violations can be that {\it (i)} the P.P. has shrunk i.e., newer versions of P.P. of an app are disclosing the collection (or sharing) of less number of PII, {\it (ii)} the behavior of app has changed i.e., newer versions of an app are leaking more PII, {\it (iii)} or both.    
We observe in Figure~\ref{fig:practices_change} that among the newer versions of P.P., the number of P.P. that had a decrease in the number of disclosed practices compared to the preceding version were in greater proportion than those that had an increased disclosure of PII.
Furthermore, as evident from Table~\ref{tab:leaksSum}, the number of leaks has also been on the rise, particularly to third-party domains. With the rise of third-party advertisements and analytics services, this comes as no surprise. The increase in the concealment of third-party domains (most recurring domains being advertisement and analytics) from the P.P. text ({\it see} Figure~\ref{fig:domsNotMent}) lends further support to this argument. 

\vspace{-0.2cm}
\section{Limitations}
\label{sec:lim}
\vspace{-0.2cm}

Despite our best efforts to longitudinally analyze apps' compliance, there are still some limitations to our work.

To scale up the study and cope with a large number of apps/versions, we leveraged automated analysis tools as well as machine learning classification. While this may result in ``false positive'' detection of privacy violations, the risk of false positives is in fact limited: static app analysis techniques are deterministic (they rely on pattern matching), and dynamic privacy leak detection has high validation accuracy (above 97\%).

The use of machine learning and natural language processing for analyzing the privacy policy texts has some limitations. Privacy policies with generic statements or subjectively redundant words are arduous to interpret without human intervention. The use of subjective verbiage in handling user's data is also not recommended by Android~\cite{googlePlayDisclosure}. 

In this study, the level of severity of privacy leaks is a fixed assumption. For example, some users may find the location as a more intrusive leak than email id and vice-versa. Furthermore, the gap between the app version release date and the P.P. date (policy snapshot that matches the release date) can be several months. This gap is because the Wayback Machine could not find a P.P. snapshot closer to the release date. Therefore, it may be possible that the policy under consideration is significantly different from the unretrievable one and thus may not be accurately reflecting the app version's data practices.      

\vspace{-0.2cm}
\section{Related Work}
\label{sec:rwork}
\vspace{-0.2cm}

	In privacy policies,  apps' developers must declare all the permissions an app will require to perform the tasks; however, previous study has shown that websites rarely disclose ``Do Not Track'' DNT in their privacy policies, and most websites do not comply with DNT even after disclosing it~\cite{libert2018www}. 
	
 	 Zimmeck et al.
 	 ~\cite{zimmeckEtAlMAPS2019}, Slavin et al.~\cite{slavin2016}, and Wang et al.~\cite{wang2018} employed static analysis for analyzing privacy policy violations in Android apps. Zimmeck et al.~\cite{zimmeckEtAlMAPS2019} and Slavin et al.~\cite{slavin2016} used Android API calls to identify the collection of users' PII from mobile devices. Using native code, Wang et al.~\cite{wang2018} check data transparency for data provided by users via GUI components. 
 	 Reyes et al.~\cite{reyes2018won}, Okoyomon et al.~\cite{okoyomon2019ridiculousness}, and Andow et al.~\cite{andow2020} leveraged dynamic analysis for identifying contradictions in the behavior of apps with the privacy policies or regulations.   
 	 
 	 Perhaps, a much closer work to our's is by Ren et al.~\cite{ren2018appversions}, where they monitored the network traffic to detect PII leaks of 512 Android apps across different versions. They show that apps leak PII to more third-party domains over time. We build on this work by conducting a compliance analysis of apps across different versions by taking the union of leaks detected in static and dynamic analysis and then comparing those leaks with the disclosure of PII in privacy policies.
	

    The above studies have significantly contributed to the understanding of privacy leaks in Android apps by performing a \textit{static} or \textit{dynamic} analysis of apps. However, these studies have been conducted at a snapshot and have not been comprehensively evaluated over time. Given that the mobile ecosystem is continuously evolving, measurement studies longitudinally starting in the present may not comprehensively illuminate privacy policy violation trend to improve compliance. To fill this research gap, our study incorporates the longitudinal analysis spanning eight years. Furthermore, static analysis alone cannot deal with dynamically loaded code and obfuscated function calls, while dynamic analysis is prone to miss part of the function calls in the app. Our methodology raises the bar in the analysis of privacy leaks by combining static and dynamic analysis in the attempt to capture most--if not all--of the leaks in Android apps. To our knowledge, the comprehensive longitudinal view on the Android app compliance with privacy policies has not been explored yet.
    
\vspace{-0.2cm}	
\section{Concluding Remarks}
\label{sec:cfwork}
\vspace{-0.2cm}

In this paper, we analyzed the 5,240 historical versions,  from 2008 to 2016,  of 268 popular Android apps and we investigated their compliance with the app privacy policies (P.P.). Our study found that most of the apps follow practices that contravene with what is declared in their privacy policy. Our results also showed that apps are becoming more prone to violating their privacy policy than before, as we observed that the percentage of released app versions that comply with their privacy policy is steadily declining over time. In particular, the newer app versions disclose fewer PII collections in their P.P. and share more of a user's private information through practices that are not mentioned in the P.P. This trend is of primary concern to users, especially considering that P.P. remain the cornerstone for protecting online privacy.  


As future work, we aim to extend our study for recent years to evaluate the change in the data disclosure and data collection (or sharing) practices caused by the GDPR. 
We also plan to study the similarity between privacy policy of non-compliant apps to determine if third party privacy policy generators have created their privacy policy. In this manner, app developers can be notified of the breach since app developers are often not experts in the domain of policy compliance and privacy laws.

\bibliographystyle{IEEEtran}
\bibliography{main}
\vspace{-0.2cm}
\section{Appendix}
\appendix
\vspace{-0.2cm}
\section{Apps Selection}
\label{sec:apps}
\vspace{-0.3cm}
For the privacy policy analysis, we obtain 3,151 privacy policies and map them to 405 apps (comprising 7,998 APKs). Among the 3,151 privacy policies, 2,455 contain at least one {\it valid} segment and are mapped to 329 apps. From the 405 apps, we successfully download and analyze (static + dynamic) 350 apps comprising 7,741 APKs (or app versions) from the unofficial Google Play API. For each APK, we obtain its release date and assign a privacy policy based on the closest timestamp after the release date. For the compliance analysis, we only consider the APKs with at least one valid segment in their assigned privacy policy. Among the 7,741 APKs (spanning 350 apps) that we analyzed, 5,240 APKs (spanning 268 apps) satisfy the criterion and are considered for compliance analysis. 
It is possible for a given app to appear in privacy policy analysis and leaks analysis, but not in compliance analysis. For example, an app has four versions (v1.1, v1.2, v1.3, and v1.4), and three unique privacy policies (pp1, pp2, and pp3). Suppose that pp1 is assigned to v1.1 and v1.2, and pp3 is assigned to v1.3 and v1.4. If pp2 contains valid segment(s), but pp1 and pp3 do not contain any valid segment, then the given app will not feature in the compliance analysis.       





\section{Label Descriptions and Classifiers Performance}
\label{sec:label_desc}
\vspace{-0.3cm}

\begin{table}[!b]
\small
\caption{\small{Descriptions of PII disclosures in the privacy policies that are collected/shared. }}
\vspace{-0.1cm}
\centering
\resizebox{.95\columnwidth}{!}{%
\begin{tabular}{@{}|l|l|@{}}
\toprule
\multicolumn{1}{|c|}{\textbf{Data Type (PII)}} & \multicolumn{1}{c|}{\textbf{Description}}                                                                      \\ \midrule
                                        
Contact                                  & Unspecified contact data.                                                   \\ \midrule
Contact\_Address\_Book                   & Contact data from a user's address book on the phone.                       \\ \midrule
Contact\_City                            & User's city.                                                            \\ \midrule
Contact\_E\_Mail\_Address                & User's e-mail.                                                          \\ \midrule
Contact\_Password                        & User's password.                                                        \\ \midrule
Contact\_Phone\_Number                   & User's phone number.                                                    \\ \midrule
Contact\_Postal\_Address                 & User's postal address.                                                  \\ \midrule
Contact\_ZIP                             & User's ZIP code.                                                        \\ \midrule
Demographic                              & User's unspecified demographic data.                                    \\ \midrule
Demographic\_Age                         & User's age (including birth date and age range).                        \\ \midrule
Demographic\_Gender                      & User's gender.                                                          \\ \midrule
Identifier                               & User's unspecified identifiers.                                         \\ \midrule
Identifier\_Ad\_ID                       & User's ad ID (such as the Google Ad ID).                                \\ \midrule
Identifier\_Cookie & User's HTTP cookies, flash cookies, pixel tags, or similar identifiers. \\ \midrule
Identifier\_Device\_ID                   & User's device ID (such as the Android ID).                              \\ \midrule
Identifier\_IMEI                         & User's IMEI (International Mobile Equipment Identity).                  \\ \midrule
Identifier\_IMSI                         & User's IMSI (International Mobile Subscriber Identity).                 \\ \midrule
Identifier\_IP\_Address                  & User's IP address.                                                      \\ \midrule
Identifier\_MAC                          & User's MAC address.                                                     \\ \midrule
Identifier\_Mobile\_Carrier              & User's mobile carrier name or other mobile carrier identifier.          \\ \midrule
Identifier\_SIM\_Serial                  & User's SIM serial number.                                               \\ \midrule
Identifier\_SSID\_BSSID                  & User's SSID or BSSID.                                                   \\ \midrule
Location                                 & User's unspecified location data.                                       \\ \midrule
Location\_Bluetooth                      & User's Bluetooth location data.                                         \\ \midrule
Location\_Cell\_Tower                    & User's cell tower location data.                                        \\ \midrule
Location\_GPS                            & User's GPS location data.                                               \\ \midrule
Location\_IP\_Address                    & User's IP location data.                                                \\ \midrule
Location\_WiFi                           & User's WiFi location data.                                              \\ \bottomrule
\end{tabular}}
\label{tab:PII_desc}
\end{table}

\begin{table}[!htb]
\vspace{-0.3cm}
\caption{\small{Performance of different classifiers. Multinomial Naive Bayes (MNB), Logistic Regression (LReg), and Linear Support Vector Classifier (SVC). The accuracy scores for SVC range from 91.16\% to 100\%. }}
\vspace{-0.1cm}
\centering
\resizebox{.8\columnwidth}{!}{%
\begin{tabular}{l|rrr|rrr}
\toprule

\multirow{2}{*}{\textbf{Classifiers}} & \multicolumn{3}{c|}{\textbf{accuracy (\%)}} & \multicolumn{3}{c|}{\textbf{precision (\%)}} \\
\cmidrule(lr){2-4} \cmidrule(lr){5-7} 
 & \textbf{MNB} & \textbf{LReg} & \textbf{SVC} & \textbf{MNB} & \textbf{LReg} & \textbf{SVC} \\

\midrule

Contact & 98.6 & 98.7 & 98.8 & 0 & 77.8 & 65.4 \\
Contact\_Address\_Book & 98.8 & 99.3 & 99.4 & 0 & 79.5 & 80.5 \\
Contact\_City & 99.6 & 99.8 & 99.8 & 0 & 81.8 & 73.3 \\
Contact\_E\_Mail\_Address & 95.9 & 97.3 & 97.2 & 84.8 & 84.8 & 83 \\
Contact\_Password & 98.8 & 99.2 & 99.4 & 0 & 70.7 & 78.6 \\
Contact\_Phone\_Number & 97 & 98.8 & 98.8 & 9.1 & 86.7 & 83.6 \\
Contact\_Postal\_Address & 98 & 98.6 & 99 & 0 & 68.5 & 80.3 \\
Contact\_ZIP & 99.4 & 99.8 & 99.8 & 0 & 94.1 & 94.4 \\
Demographic & 98.7 & 99.6 & 99.6 & 25 & 79.2 & 78.9 \\
Demographic\_Age & 98.3 & 99.3 & 99.4 & 16.7 & 89.3 & 90 \\
Demographic\_Gender & 98.9 & 99.6 & 99.6 & 16.7 & 79.3 & 79.3 \\
Identifier & 99.1 & 99.1 & 99.1 & 0 & 0 & 66.7 \\
Identifier\_Ad\_ID & 98.8 & 99.7 & 99.8 & 0 & 97.7 & 97.8 \\
Identifier\_Cookie & 95.4 & 98.7 & 98.5 & 90.3 & 88 & 88.6 \\
Identifier\_Device\_ID & 97.9 & 99 & 99 & 73.6 & 86.7 & 83.7 \\
Identifier\_IMEI & 99.4 & 99.8 & 99.9 & 50 & 92.6 & 93.1 \\
Identifier\_IMSI & 99.9 & 99.9 & 99.9 & 0 & 100 & 100 \\
Identifier\_IP\_Address & 99.1 & 99.4 & 99.4 & 90.1 & 94.1 & 94.8 \\
Identifier\_MAC & 99.3 & 99.9 & 99.9 & 0 & 92.6 & 96.3 \\
Identifier\_Mobile\_Carrier & 99.4 & 99.7 & 99.7 & 0 & 100 & 85.7 \\
Identifier\_SIM\_Serial & 99.8 & 99.8 & 99.8 & 0 & 44.4 & 56.3 \\
Identifier\_SSID\_BSSID & 99.8 & 99.8 & 99.9 & 0 & 0 & 100 \\
Location & 96.7 & 98.8 & 98.8 & 80.9 & 86.5 & 87 \\
Location\_Bluetooth & 99.4 & 99.8 & 99.8 & 0 & 80 & 81 \\
Location\_Cell\_Tower & 99.5 & 99.9 & 99.9 & 25 & 79.2 & 81.8 \\
Location\_GPS & 99 & 99.7 & 99.7 & 64.7 & 90.2 & 88.7 \\
Location\_IP\_Address & 99.3 & 99.6 & 99.6 & 0 & 85.7 & 78.3 \\
Location\_WiFi & 99.6 & 99.8 & 99.8 & 70 & 77.8 & 83.3 \\
Performed & 90.5 & 91.7 & 91.2 & 91.7 & 90.9 & 86.3 \\
Not\_Performed & 98.6 & 98.9 & 98.7 & 3.2 & 34.2 & 30.5 \\
1stParty & 91 & 92.5 & 92.3 & 81 & 85 & 82.7 \\
3rdParty & 94.9 & 96.7 & 96.7 & 53.3 & 73.8 & 71.6 \\

\midrule
average & 98.06 & 98.82 & \cellcolor[HTML]{6ddf69}98.82 & 28.94 & 77.22 & \cellcolor[HTML]{6ddf69}81.92 \\ 
\bottomrule
\end{tabular}}
\label{tab:classifiers}
\vspace{-0.2cm}
\end{table}

Table~\ref{tab:PII_desc} describes the various PII labels used in our privacy policy classification. Table~\ref{tab:classifiers} shows the performance of various machine learning classification algorithms on the test data-set of APP-350 corpus. 
\end{document}